
\magnification=1200
\baselineskip=16pt
\vskip 2cm
\centerline{\bf Correlations in Random Ising Chains at Zero Temperature}
\vskip 2cm
\centerline{Ferenc Igl\'oi}
\vskip 1cm
\centerline{Institut f\"ur Theoretische Physik, Universit\"at zu K\"oln}
\centerline{Z\"ulpicherstr. 77, D-50937 K\"oln, FRG}
\centerline{and}
\centerline{Research Institute for Solid State Physics}
\centerline{H-1525 Budapest, P.O.Box 49, Hungary \footnote{$^{\dag}$}{Permanent
and
present address}}
\vskip 2cm
{\bf Abstract:}
We present a general method to calculate the connected correlation function
of random Ising chains at zero temperature. This quantity is shown to
relate to the surviving probability of some one-dimensional, adsorbing
random walker on a finite intervall, the size of which is controlled
by the strength of the randomness. For different random field and random
bond distributions the correlation length is exactly
calculated.
\vskip 3cm
PACS-numbers: 05.50.+q; 75.10.Nr
\vskip 1cm
Short title: {\it Random Ising Chains}
\vfill
\eject
{\bf 1 Introduction:}
\bigskip
The perfect crystal is the result of physical abstraction
in real materials one should always reckon on the existence of impurities
and different types of lattice defects. There are important problems in the
field of random systems such as localization in a disordered medium, the
spin glass behaviour, diluted magnets etc. In order to obtain a theoretical
understanding of these phenomena magnetic models with quenched disorder
have been introduced and studied by different methods
(for recent reviews see Refs.[1-3]). The random magnetic models have
unusual low-temperature properties. These are a consequence of the
complex structure of low energy
metastable states generated by quenched disorder
and frustration, which are beleived the main ingredients of spin-glass
behaviour[4]. The simplest
system in this field is the random Ising model, which can be experimentally
realised as diluted antiferromagnets[5]. The model
already in one dimension shows interesting features, although its physical
quantities are singular only at zero temperature.

The one-dimensional random Ising model - inspite
of its low dimension - is exactly soluble only for a few specific form of
randomness. One of those is the random bond Ising chain in a uniform
field (RBIM) defined by the Hamiltonian:
$${\cal H}_{RB}=-\sum J_{i,i+1} \sigma_i \sigma_{i+1} - h \sum \sigma_i
\eqno(1)$$
Here $\sigma_i=\pm 1$ and the exchange integral $J_{i,i+1}$ is equal to
$J>0$, with probability $p$ and $-J$ with probability $q=1-p$, and the
bond disorder is quenched. For this model the ground state energy, the
zero point entropy and magnetization have been calculated by Derrida and
co-workers[6] (see also in Ref[7]).

Another type of random system is the ferromagnetic Ising model in a
random field (RFIM) with the Hamiltonian:
$${\cal H}_{RF}=-J \sum  \sigma_i \sigma_{i+1} - \sum h_i \sigma_i
\eqno(2)$$
At $T=0$ the thermodynamic properties of this model have been
calculated for the binary distribution
of the random fields[6]:
$$P(h)=p \delta(h-H) + q \delta(h+H) \eqno(3)$$
as well as for the asymmetric binary distribution[8]:
$$P(h)={1 \over 2} \delta(h-H_1-H_0) + {1 \over 2} \delta(h+H_1-H_0) \eqno(4)$$
The first exact results at $T \ne 0$ have been obtained by Grinstein and
Mukamel[9] with the diluted symmetric binary distribution:
$$P(h)={p\over 2}[ \delta(h-H) + \delta(h+H)] + q \delta(h) \eqno(5)$$
in the limit $H \to \infty$. Very recently the non-linear and higher order
susceptibilities of the model have also been determined[10].

A series of RFIM-s with continuous random field distribution has been
studied by Luck and Nieuwenhuizen[11-13] at arbitrary temperatures. One
of these models is characterised by the diluted symmetric exponential
distribution $R(x)dx$ as:
$$R(x)={p \over 2} e^{-x}+q\delta(x) \eqno(6)$$
with $h_i=\tilde H x_i$, $\tilde H >0$ and
$-\infty < x_i < \infty$.

As far as the correlation functions of random Ising models are concerned
only a few results are available. Derrida[14] has pointed out that the
spin-spin correlation function is not a self-averaging quantity, therefore
its average differs from the most probable value[15]. Exact results in the
whole temperature range are available for the Grinstein-Mukamel model[9]
as well as for the diluted symmetric exponential distribution in eq(6)[12].
Most recently Farhi and Gutmann[16] have calculated the zero temperature
correlation function of the RFIM with the binary distribution in eq(3).
We also mention a related exact study of the pair-correlation function
of a one-dimensional lattice gas model in a random potential at zero
temperature[17].

In this paper we study the connected correlation function of random
Ising chains defined as:
$$\chi(l)=\left[ <\sigma_i \sigma_{i+l}>-<\sigma_i>< \sigma_{i+l}>
\right]_{av} \eqno(7)$$
where $<...>$ denotes the thermodynamic average and $[...]_{av}$ stands
for the quenched average over the random variables. The thermodynamic
average in eq(7) could be non-zero even at $T=0$ provided the ground state
of the system is highly degenerate. This can be seen in two-dimensional
frustrated models without randomness, in which the zero temperature
correlations either decay as a power low[18-21] or exponentially[22].

For random Ising chains the correlation function in
eq(7) is usually calculated in the transfer matrix formalism[9,12]. At
$T=0$, however, one may use another approach based on an analysis of the
degenerate ground state configurations. For a given quenched disorder,
due to frustration, there are spins in the system which are "loose", i.e.
they are free to point in any direction. These loose spins may form a
domain, and the connected correlation function is non-zero, only if both
spins considered belong to the same domain of loose spins. At this point
calculation of the correlation function at $T=0$ is essentially reduced
to an investigation of the size distribution of domains of loose spins,
which in turn is equivalent to a one-dimensional random walker problem
on a finite intervall.

The setup of the paper is the following. In Section 2 we present our method
to calculate the connected correlation function for random Ising chains
at $T=0$. In Section 3 and 4 correlation lengths are calculated for RFIM and
RBIM, respectively. Finally, the results are discussed in Section 5.
\vskip 1cm
{\bf 2 Connected correlations and their relation with random walkers}
\bigskip
In this Section we develope a formalism to calculate the connected correlation
function for random field Ising models at $T=0$. It will be shown in Section 4
how these results can be applied for the random bond problem.

The random fields we consider have a discrete distribution, furthermore the
possible values
of $h_i$ are integer multiples of a unit, denoted by $H$, i.e.
$h_i=m \times H$. For the binary distribution in eq(3) $m=\pm 1$, while
for the diluted symmetric distribution in eq(5) $m=0,\pm 1$. Continuous
distributions, like in eq(6) can also be discretised, an example is shown
in Section 3.3. In this way one can also treat the asymmetric binary
distribution in eq(4)
provided the ratio of the parameters $H_1$ and $H_0$ is rational.

The structure of ground state configurations of a RFIM with discrete randomness
is thoroughly analysed in the literature (see c.f. in Refs[2,3]). In the
weak-coupling limit, when $2J<min\{|h_i|\}=h_{min}$ the spins are frozen to
the direction of the local fields. For stronger couplings, so that $2J >
h_{min}$
there is a tendency for neighbouring spins to align parallel with each other,
so that domains of parallel spins are formed. With increasing value of the
coupling the average size of a domain increases, but the ground state never
consists of one single domain. It can be understood, since the necessary
energy to create a domain wall - $2J$ - can be accumulated from fluctuations
of the random field, even if the local field is arbitrarily small.

The size of random field fluctuations is characterised by the integrated
random field function, defined as: $H(k)=\sum_{i=1}^k h_i$. For an illustration
we draw this function on Fig 1 for a given random field distribution together
with the corresponding ground state configuration of the system at some
value of the coupling $J$. As is seen on this Figure the first domain wall
is located between spins 3 and 4, at a local maximum of $H(k)$. Indeed, the
sum of random field energies $H(6)-H(3)<-2J$ covers the cost of creation a
domain wall.

The position of the second domain wall, however is not unique: it can be at
any of the three degenerate local minima located at $(6,7)$, $9,10)$ or
$(11,12)$. Energetically there is no difference between these
configurations, since the corresponding random field energies are the
same: $H(6)=H(9)=H(11)$. As a consequence in the intervall $\{ 7,11\}$
the position of the spins in the ground state is not fixed. Such a region
will be called as a domain of "loose" spins (DLS).

Based on this example we can easily postulate the properties of a DLS.
First let us restrict ourselves to a DLS which separates a $\downarrow ...
\downarrow$ and a $\uparrow ... \uparrow$ domain, as shown in our
example on Fig 1. Such a DLS is bounded by two degenerate local minima. Inside
the DLS the integrated random field function relative to its value at the
boundaries - $\Delta H(k)$ - does not exceed $2J$, thus the energy necessary
to create a domain wall can not be accumulated from the random field. On the
other hand in both directions outside of the DLS $\Delta H(k)$ exceeds $2J$,
before crossing zero. This last condition ensures the existence of
$\downarrow ...
\downarrow$ and $\uparrow ... \uparrow$ ferromagnetic domains at two sides
of the DLS.

The other type of DLS separating $\uparrow ... \uparrow$ and $\downarrow ...
\downarrow$ domains can be characterised similarly. In this case after the
transformation $\sigma_i \to -\sigma_i$, $h_i \to -h_i$ the previous
considerations can be applied for $-H(k)$. We mention, if $2J/H$ is an integer
the degeneracy of the ground state is higher, than for a slightly
larger or smaller value of the coupling. It is connected to the fact that in
this case conditions both for $\uparrow \downarrow$ and $\downarrow \uparrow$
DLS can be satisfied at the same time. In the following we are not going to
deal with such situations, thus our considerations apply for $2J/H \ne
integer$.

Since position of spins in a DLS is not fixed these regions are the source of
non-zero ground state entropy in a RFIM. These regions are also responsible
for non-vanishing value of the connected correlation functions at $T=0$. It is
easy to see that the thermal average of $\chi(l)$ in eq(7), which is now
performed over the degenerate ground state configurations, is non-zero
only if the two spins are in the same DLS. If one spin is in a ferromagnetic
domain its value is the same for all ground state configurations, consequently
the connected correlations are zero. On the other hand if two spins are in
different DLS they are
independent variables, thus again the connected correlations are vanishing.

The above relationship between $\chi(l)$ and DLS makes it possible to
calculate the leading behaviour of the connected correlation function in a
RFIM in a simple way. To do this one should i) first consider DLS regions of
length $\tilde l >l$ and determine the $W(\tilde l)$ probability that a point
of the line belongs to one of those DLS. ii) Then find the probability that the
other
reference point of $\chi(l)$ is also on the same DLS and iii) finally calculate
the thermal and quenched averages in eq(7) with the condition that both
endpoints of $\chi(l)$ are on the same DLS. In the following we show that the
leading behaviour of $\chi(l)$ is determined by the probability $W(l)$, which
has an exponential dependence on $l$, whereas the probabilities indicated in
ii) and iii) are comperatively negligable, since they depend on $l$ as
power laws.

To calculate the probability $W(\tilde l)$ one can use a geometrical
interpretation of a DLS as a one-dimensional random walker with steps $h_i/H$
on an intervall consisting of
$$L=\left[{2J \over H} \right]+1 \eqno(8)$$
points. Here $[x]$ denotes the integer part of $x$ and $x$ is non-integer.
The walker starts at one endpoint of the intervall and after $\tilde l$ steps
made on the strip returns to the same endpoint. For large $\tilde l$ the
leading $\tilde l$ dependence of $W(\tilde l)$ is exponential, and
$W(\tilde l)$ corresponds to the surviving probability of the adsorbing
random walker:
$$W(\tilde l) \sim \exp \left[-\tilde l/\xi(L)\right] \eqno(9)$$

Next we show that the probabilities mentioned in ii) and iii) have weaker
$l$ dependence than the surviving probability. It is clear from a simple
geometrical consideration that the probability in ii) is at most $\sim 1/l$.
On the other hand to estimate the conditional probability in iii) one should
first notice that the different ground states are characterised by one
parameter, the position of the domain wall. Thus an average over this parameter
is equivalent to the thermal end quenched averages in eq(7). Using the fact
that the number of possible positions of a domain wall in a DLS of length $l$
is proportional to $l$ one can estimate the conditional probability in iii)
as $\sim l^{-2}$.

The connected correlation function $\chi(l)$ is then obtained by summing the
probabilities for $\tilde l \ge l$ with the result in leading order:
$$\chi(l) \sim W(l) \sim \exp \left[-l/\xi(L)\right] \eqno(10)$$
where $L$ is given in eq(8).
In the following Section the correlation length $\xi(L)$ will be calculated for
different random field distributions.
\vskip 1cm
{\bf 3 Random field Ising models}
\bigskip
The surviving probability in eq(10) can be most easily calculated in the
transfer matrix formalism[23]. In this case the elements of the transfer
matrix $T(n,m)$, $n,m=1,2,...,L$ are given as the probability of a step from
position $m$ to $n$. In the RFIM lanquage $T(n,m)=P(h(n,m))$, where
$h(n,m)=(n-m)\times H$. A matrix-element of $T$ is zero, whenever the
corresponding $h(n,m)$ is not contained in the set of random fields
of the model. The leading eigenvalue of the transfer matrix - $\lambda(L)$ -
is connected to the surviving probability as:
$$W(l) \sim \lambda(L)^l \eqno(11)$$
thus the correlation length in eq(10) is given by
$$\xi(L)=-{1 \over \log \lambda(L)} \eqno(12)$$
For a general RFIM simple analytical results can be obtained for $2J/H<1$
and in the strong coupling limit $2J/H \gg 1$. In the former case $L=1$,
the intervall of the walker consists of one single point:
$$\lambda(1)=P(0)~~~~2J/H<1 \eqno(13)$$
Therefore non-vanishing correlations can only be present in diluted models (see
cf.
eqs.(5) and (6)). In the strong coupling limit, which corresponds to $L \gg 1$
we consider distributions with zero average $<h_i>=0$. Then the finite size
corrections to the leading eigenvalue are quadratic:
$$1-\lambda(L) \sim L^{-2} \sim (H/J)^2~~~~H/J \ll 1 \eqno(14)$$
which follows from the Gaussian nature of the free random walk[23]. According
to eqs.(12) and (14)
$$\xi(J/H) \sim (J/H)^2~~~~J/H \gg 1 \eqno(15)$$
For intermediate (non-integer) values of $2J/H$ the leading eigenvalue
of the transfer matrix can be calculated numerically, so that - together
with the asymptotic relation in eqs(15) - one can in principle obtain the
correlation length of connected correlations for all types of RFIM-s.
In the following we present three examples, in which the calculation can be
performed analytically for all non-integer values of $2J/H$.
\vfill
\eject
{\bf 3.1 Binary distribution}
\bigskip
The transfer matrix corresponding to the distribution in eq(3) is tridiagonal
and given as:
$$T_1=\pmatrix{0&p&&&&\cr
               q&0&p&&&\cr
                &q&0&p&&\cr
                &&&\ddots&&\cr
                &&&&&p\cr
                &&&&q&0\cr} \eqno(16)$$
The (non-normalized) leading right eigenvector of this matrix is
$$\Phi_1(l)=\left( q \over p \right)^{l/2} \sin \left(l {\pi \over L+1}
\right) \eqno(17)$$
and the corresponding leading eigenvalue
$$\lambda_1(L)=2 \sqrt{pq} \cos\left( {\pi \over L+1}\right) \eqno(18)$$
The correlation length is then can be obtained from eqs(12) and (8). This
result agrees with that of Ref[16].
\bigskip
{\bf 3.2 Diluted symmetric binary distribution}
\bigskip
The transfer matrix corresponding to the distribution in eq(5) is symmetric
and tridiagonal:
$$T_2=\pmatrix{q&p/2&&&&\cr
               p/2&q&p/2&&&\cr
                &p/2&q&p/2&&\cr
                &&&\ddots&&\cr
                &&&&&p/2\cr
                &&&&p/2&q\cr} \eqno(19)$$
The leading eigenvector of $T_2$ is the same as in eq(17), however with
$q/p=1$ and the leading eigenvalue is given as:
$$\lambda_2(L)=q+p \cos \left( {\pi \over L+1} \right) \eqno(20)$$
For $L=1$, i.e. $2J/H<1$, $\lambda_2(1)=q$ and the general relation in eq(13)
is recovered, which agrees also with the result of Grinstein and Mukamel[9]
as $2J/H \to 0$.
\bigskip
{\bf 3.3 Diluted symmetric exponential distribution}
\bigskip
We consider the discretised version of the distribution in eq(6) when the
allowed values of the random field are $h_i=H\times i$, $i=0,\pm 1,\pm 2,...$
and the probability distribution is given as:
$$P(h)=c \exp\left(-\alpha \left|{h \over H}\right| \right) \delta(h-Hi) +
c_0 \delta(h) \eqno(21)$$
Here $\alpha>0$ and the distribution is normalised with $c=p/2(\exp \alpha-1)$,
$c_0=q-c$ and $p+q=1$. Taking the limit
$$\alpha \to 0^{+},~~~H \to 0,~~~\tilde H=H/(\exp \alpha -1)= {\rm finite}
\eqno(22)$$
one arrives to a continuum description in the variables $h_i=\tilde H x_i$,
$-\infty < x_i < \infty$, with the probability distribution $R(x)dx$ given in
eq(6).

We write the transfer matrix of the problem in terms of the variables
$\omega=\exp(-\alpha)$, $\omega_0=q$ and $\kappa=c/q$ as:
$$T_3=\omega_0
\pmatrix{1&\kappa\omega&\kappa \omega^2&\ldots&\kappa \omega^{L-1}\cr
              \kappa\omega&1&\kappa \omega&&\cr
               \kappa\omega^2&\kappa \omega&1&&\cr
               \vdots&&&\ddots&\cr
               \kappa \omega^{L-1}&\ldots&&&1\cr} \eqno(23)$$
We mention that the same transfer matrix belongs to a directed polymer[23]
on a strip of width $L$ on the square lattice. In this case $\omega_0$ and
$\omega$ are the
monomer fugacities for steps along and perpendicular to the strip,
respectively and $\kappa$ denotes the statistical weight corresponding
to a turn of the chain by $90^{o}$.

The eigenvalue problem of $T_3$ in eq(23) is similar to that of the
unrestricted directed polymer[24]. The leading eigenvector is given as
$$\Phi_3(l)=\cos \left[ \left(l-{L+1 \over 2} \right) \phi \right] \eqno(24)$$
where $\phi$ is the smallest root of the equation:
$${\rm cot}\left({(L-1)\phi \over 2} \right)={\sin \phi \over
\cos \phi - \omega} \eqno(25)$$
The leading eigenvalue is then:
$$\lambda_3(L)={(1-\omega^2)\kappa \omega_0 \over 1-2 \omega \cos \phi +
\omega^2}+(1-\kappa) \omega_0 \eqno(26)$$
The correlation length can be obtained from eq(12) using
the correspondences between $\omega, \omega_0, \kappa$ and the original
parameters of the distribution in eq(21).

Now we evaluate our results in eqs(25) and (26) in the continuum limit
of eq(22), which reads as $\omega \to 1^-$. The smallest
root of eq(25) is proportional to $1-\omega$, so that written it in the form
of $\phi=\gamma (1-\omega)$ one gets the following relation from eq(25):
$${\rm tan}^{-1}\left( {1 \over \gamma} \right)={J \over \tilde H} \gamma
\eqno(27)$$
Then the leading eigenvalue is given by
$$\lambda_3={1+q \gamma^2 \over 1+ \gamma^2} \eqno(28)$$
in the continuum limit. These results are identical to those obtained by
Luck and Nieuwenhuizen[12] using the continuous distribution in eq(6) and
taking the non-trivial limit $T \to 0^+$. We note that the correlation
function calculated with the continuous distribution is
discontinuous at $T=0$, thus the limits $H \to 0$ and $T \to 0$ can not
be interchanged.
\vskip 1cm
{\bf 4 Random bond Ising model}
\bigskip
The formalism developed in Section 2 can also be applied to calculate
$\chi(l)$ for the RBIM with the Hamiltonian in eq(1). It is easy to see
that after the transformation $\sigma_i \to \sigma_i h/h_i$
in eq(2) one arrives to a RBIM with $J_{i,i+1}=J/h_i h_{i+1}$. The inverse
transformation is given by
$$\sigma_i \to \sigma_i \prod^{i-1}_{k=1}\left( {J_{k,k+1} \over J} \right)
\eqno(29)$$
thus the random fields can be expressed through the random bonds as
$$h_i=h \prod^{i-1}_{k=1}\left( {J_{k,k+1} \over J} \right)
\eqno(30)$$
The random fields in eq(30) can take the values $\pm h$ like to the binary
distribution
in eq(3), however these $h_i$-s are correlated in different sites, since
$$<h_i h_{i+n}>=(p-q)^n \eqno(31)$$
Thus the symmetric distribution with $p=q=1/2$ is exceptional, in which case
the RBIM is equivalent to a RFIM with
the symmetric binary distribution in eq(3).

For $p \ne q$ the probability is connected with bond variables, thus with
the sign of the product of two consecutive random fields:
$$P(h_{i-1} h_i)=\cases{p&$h_{i-1}h_i>0$\cr
                        q&$h_{i-1}h_i<0$\cr} \eqno(32)$$
To write down the transfer matrix of this problem we work with the
bond variable $\bar H(i)=(H(i-1)+H(i))/2$, which may take $L-1$ different
values: $1/2,3/2,...,L-1/2$. The one-step transfer matrix
$T_{RB}(i,i+1)$ is
different if the number of the step $i$ is even or odd. Therefore one should
consider the two-step transfer matrix defined as:
$$T_{RB}(i,i+2)=
\pmatrix{q^2&qp&p^2&&&&&\cr
         qp&q^2&qp&&&&&\cr
         &qp&q^2&qp&p^2&&&\cr
         &p^2&qp&q^2&qp&&&\cr
         &&&qp&q^2&qp&p^2&\cr
         &&&p^2&qp&q^2&qp&\cr
         &&&&&&\ddots&\cr
         &&&&&&&\ddots\cr} \eqno(33)$$
The transfer matrix connecting the states in steps $i+1$ and $i+3$ is the
transpose of $T_{RB}(i,i+2)$, thus both have the same eigenvalue spectrum.
Writing the leading eigenvalue as $\lambda_{RB}^2$
the correlation length is then obtained from the logarithm of
$\lambda_{RB}$ through eq(12).
Since the transfer matrix in eq(33) is non-symmetric and non-tridiagonal
we could solve its eigenvalue problem analytically only in a few special
cases.

In the limit $p \to 0$ the transfer matrix is tridiagonal in linear order
of $p$ and can be solved by the same eigenvector as $T_2$ in eq(19). The
leading eigenvalue is given as
$$\lambda_{RB}=1-{p \over 2}\left(1-\cos {\pi \over L}\right) + O(p^2)
\eqno(34)$$
In the symmetric distribution $p=q=1/2$ the right eigenvector is given by
$$\Phi_{RB}(2l)=\Phi_{RB}(2l+1)=\sin\left[ (2l+1){\pi \over L+1} \right]
\eqno(35)$$
for $l=0,1,...,L/2-1$. The corresponding eigenvalue
$$\lambda_{RB}=\cos\left( \pi \over L+1 \right)~~~~(p=q=1/2) \eqno(36)$$
is the same as for the RFIM with symmetric distribution in eq(18), which is
in accord with our previous claim about the equivalence of the two problems.

Finally, we consider the limit $q \to 0$. Then the eigenvector is given
in leading order as:
$$\eqalign{\Psi(2l)=\Psi(L-2l+1)=q^{l-1 \over L-1}\cr
\Psi(2l-1)=\Psi(L-2l+2)=q^{1-{l \over L-1}}\cr} \eqno(37)$$
for $l=1,2,...,(L-1)/2$. The corresponding eigenvalue:
$$\lambda_{RB}=q^{1/L}~~~~q\ll 1 \eqno(38)$$
Analysing the $p$-dependence of the correlation length one can see that
it starts with zero in the ferromagnetic limit $p \to 1$, stays finite for
non-zero concentration of ferromagnetic bonds and finally diverges in the
antiferromagnetic limit $p \to 0$.
\vskip 1cm
{\bf 5 Discussion}
\bigskip
In this paper the connected correlation function of random Ising chains
is investigated at zero temperature. Our study is based on an analysis of the
ground state configurations of these systems. Due to frustration there
are spins in the ground state which are "loose", i.e. their position
is not fixed by the interaction and the external field. These "loose"
spins form domains, and between two spins in the same domain
there are non-vanishing correlations. It was shown then that the connected
correlation function $\chi(l)$ is primerely determined by the probability
of having a DLS of size $l$ in the system. Finally, this probability was
calculated using an analogy with the surviving probability of some random
walker on a finite intervall.

Considering different types of random field and random bond distributions
we have calculated the correlation length using the transfer matrix method.
Analysing these results one my observe two different
behaviour in the weak disorder limit. The correlation length either
vanishes as $\xi \sim 1/ \log(1/p)$ or it is divergent like $\xi \sim 1/p$.
The former behaviour is found in the RFIM with binary distribution in eq(18)
as well as for the RBIM in the pure ferromagnetic limit $q \to 0$ in eq(38).
On the other hand the correlation length is diverging in the diluted
RFIM in eqs(20) and (28) as $p \to 0$. Similar $p$ dependence is observed
in the RBIM in the antiferromagnetic limit, i.e. when $p \to 0$ in eq(34).

The exponential decay of correlations for random Ising chains is found as
a general rule. One may find, however, a slower decay of correlations
if the strength of
randomness is smoothly position dependent. Let us consider a semi-infinite
RFIM in which the strength of the random field decays like $h_i \sim i^{-s}$
from the surface. Then the equivalent random walker has to be considered on
an intervall the size of which is increasing in time. Using results
about random walkers[25] and directed polymers[24] inside a parabola,
one can say that the decay of correlations in this inhomogeneuosly
disordered system is of a stretched exponential form for $0<s<2$, whereas
it can be described as a power law for $s \ge 2$.
\vskip 1cm
{\bf Acknowledgement}: The author thanks for the Institute f\"ur Theoretische
Physik, Universit\"at zu K\"oln for hospitality and the
Sonderforschungsbereich 341 K\"oln-Aachen-J\"ulich for financial support.
This work has been supported by the Hungarian National Research Fund
under Grant No. OTKA T012830.
The author is indebted to U. Behn, M. Schreckenberg and L.H. Tang for
interesting discussions.
\vfill
\eject
{\bf References:}
\vskip 1cm
\item{[ 1]} T. Nattermann and J. Villain, Phase Transitions 11, 5 (1988)
\bigskip
\item{[ 2]} T. Nattermann and P. Rujan, Int. J. Mod. Phys. B3, 1597 (1989)
\bigskip
\item{[ 3]} J.M. Luck, {\it Syst\'emes D\'esordonn\'es Unidimensionels}
(Collection Al\'ea-Saclay, 1992)
\bigskip
\item{[ 4]} G. Toulouse, Commun. Phys. 2, 115 (1977)
\bigskip
\item{[ 5]} S. Fishman and A. Aharony, J. Phys. C12, L729 (1979)
\bigskip
\item{[ 6]} B. Derrida, J. Vannimenus and Y. Pomeau, J. Phys. C11, 4749 (1978)
\bigskip
\item{[ 7]} J.K. Williams, J. Phys. C14, 4095 (1981)
\bigskip
\item{[ 8]} U. Behn, V.B. Priezzhev and A. Zagrebnov, Physica A167, 481 (1990)
\bigskip
\item{[ 9]} G. Grinstein and D. Mukamel, Phys. Rev. B27, 4503 (1983)
\bigskip
\item{[10]} E. Amic and J.M. Luck, Saclay preprint SPhT/94/005
 \bigskip
\item{[11]} Th.M. Nieuwenhuizen and J.M. Luck, J. Phys. A19, 1207 (1986)
\bigskip
\item{[12]} J.M. Luck and Th.M. Nieuwenhuizen, J. Phys. A22, 2151 (1989)
\bigskip
\item{[13]} J.M. Luck, M. Funke and Th. M. Nieuwenhuizen, J. Phys. A24, 4155
(1991)
\bigskip
\item{[14]} B. Derrida, in {\it Les Houches Winter School Proceedings} (1983)
\bigskip
\item{[15]} B. Derrida and H.J. Hilhorst, J. Phys. C14, L539 (1981)
\bigskip
\item{[16]} E. Farhi and S. Gutmann, Phys. Rev. B48, 9508 (1993)
\bigskip
\item{[17]} Y. Fonk and H.J. Hilhorst, J. Stat. Phys. 49, 1235 (1987)
\bigskip
\item{[18]} J. Stephenson, J. Math. Phys. A11, 413 (1970)
\bigskip
\item{[19]} G. Forgacs, Phys. Rev. B22, 4473 (1980)
\bigskip
\item{[20]} I. Peschel, Z. Phys. B45, 339 (1982)
\bigskip
\item{[21]} W.F. Wolff and J. Zittartz, Z. Phys. B47, 341 (1982)
\bigskip
\item{[22]} A. S\"ut\H o, Z. Phys. B44, 121 (1981)
\bigskip
\item{[23]} V. Privman and N.M. \v Svraki\'c,{\it Directed Models of
Polymers, Interfaces and Finite-Size Properties}, Lecture Notes in Physics
Vol. 338 (Springer, Berlin, 1989)
\bigskip
\item{[24]} F. Igl\'oi, Phys. Rev. A45, 7024 (1992)
\bigskip
\item{[25]} L. Turban, J. Phys. A25, L127 (1992)
\vfill
\eject
{\bf Figure Caption:}
\vskip 1cm
\item{Fig.1.} The integrated random field function $H(i)$ for some fixed
values of the random field. The corresponding ground state of the system
at a given value of the coupling $J$ is indicated below. Here vertical
dashed lines denote the possible domain wall positions. State of spins
in the region $\{7,11\}$ is not fixed, they form a DLS(see text).
\vfill
\eject
The integrated random field function $H(i)$ for some fixed
values of the random field. The corresponding ground state of the system
at a given value of the coupling $J$ is indicated below. Here vertical
dashed lines denote the possible domain wall positions. State of spins
in the region $\{7,11\}$ is not fixed, they form a DLS(see text).
\vfill
\eject
\bye